\documentclass[a4paper,12pt,english,german]{article}

\usepackage{mathtools}
\usepackage{graphicx}

\begin{document}

{\bf Kraus operators for a pair of interacting qubits: a case study}

\bigskip

{  Momir Arsenijevi\' c$^a$, Jasmina Jekni\' c-Dugi\' c$^b$, Miroljub Dugi\' c$^a$}

\bigskip

$^a$University of Kragujevac, Faculty of Science, Radoja
Domanovi\' ca 12, 34000 Kragujevac, Serbia

$^b$University of Ni\v s, Faculty of Science and Mathematics, Vi\v
segradska 33, 18000 Ni\v s, Serbia

\bigskip

{\bf Abstract}
The Kraus form of the completely positive dynamical maps is appealing from the mathematical and the point of the diverse applications of the open quantum systems theory. Unfortunately, the Kraus operators are poorly known for the two-qubit processes. In this paper, we derive the Kraus operators for a pair of interacting qubit, while the strength of the interaction is arbitrary. One of the qubits is subjected to the x-projection spin measurement. The obtained results are applied to calculate the dynamics of the initial entanglement in the qubits system. We obtain the loss of the correlations in the finite time interval; the stronger the inter-qubit interaction, the longer lasting entanglement in the system.

\bigskip

{\bf 1. Introduction}

The "integral", i.e. so-called, Kraus form [1] of a completely positive dynamical map for an open quantum system [2, 3]
is appealing for the mathematical reasons.  Mathematical existence of the Kraus form for such processes is guaranteed by the Kraus theorem,
universally [1-3]. On the other hand, a Kraus-form (KF) may be regarded as a solution to a differential
master equation (ME) for the open system's statistical operator (density matrix); a case when no ME exists for the process can be found e.g. in Refs. [4,5].

The Kraus operators are often  constructed
due to some physical assumptions or understanding of the underlying physical processes [6]. Nevertheless, such derivations may not provide the full physical (e.g. microscopic) details [7].
One way to obtain a proper KF for the open system's dynamics is derivation from the related master equation for the process [7,8]--if such an ME exists [4,5].
To this end, it is important to note: phenomenological derivations of MEs may also be unreliable--often there appear certain subtleties of both mathematical and physical nature
as well as unexpected pitfalls [9,10].

Having this in mind as well as the above-distinguished usefulness of KF, in this paper we derive the Kraus operators
starting from a microscopically derived master equation for a pair of two-level systems (qubits). We are concerned with an ancilla qubit interacting with another qubit, which is subjected to a
$S_x$ spin-projection quantum measurement. Usefulness of the KF for the process is emphasized by application of our results in investigating the dynamics of entanglement in the
qubits system.

{\bf 2. The master equation and the task}
\bigskip

The total, isolated, system is described by the Hamiltonian:

\begin{equation}
H = H_{1\circ} + H_{2\circ} + H_{E_1\circ} + H_{12} + H_{1E_1} ,
\end{equation}

\noindent where the symbol "$\circ"$ stands for the subsystems self-Hamiltonians and the rest are the interaction terms.
While the self-Hamiltonian terms are standard (see below), the qubits interaction
is chosen [6,11]:

\begin{equation}
H_{12} = \beta S_{1z} \otimes S_{2z},
\end{equation}

\noindent where the 1/2-spin operators $S_{p  z} = \sigma_{p z} /2, p = 1,2$ and we take $\hbar =1$, while the interaction with the environment:

\begin{equation}
H_{1 E_{1}} =S_{1x} \otimes \int_0^{\nu_{max}} d\nu h(\nu) (a^{\dag}_{\nu} + a_{\nu} ) \equiv  S_{1x} \otimes B_{E_1},
\end{equation}

\noindent where appear the annihilation and creation operators satisfying the standard Bose-Einstein commutation $[a^{\dag}_{\nu}, a_{\nu'} ]
= - \delta (\nu - \nu') $.

For non-interacting qubits, i.e. for $\beta = 0$,
the qubits can be described by mutually independent dynamics.  However, for the interacting qubits ($\beta \neq 0$), dynamics of the qubits cannot be mutually independent.
Therefore we regard the pair of qubits, $1+2$, as an open
system subjected to the  environment $E_1$ as described above.

Physically, eq.(3) describes a quantum measurement of the $S_{1x}$ observable. We assume the initial tensor product state $\rho_{12}\otimes \rho_{E_1}$
and the  weak coupling limit for eq.(3), while the environemnt $E_1$ being in the thermal state $\rho_{E_1}=\rho_{th}=\exp[-H_{E_1\circ}/k_B T]/Z$ on temperature $T$; $k_B$
is the Boltzmann constant and $Z$ is the normalization "statistical sum". That is, we consider the time-homogeneous, completely positive and trace preserving process for the $1+2$ system
and below we derive the proper master equation in the weak-coupling limit.
Such physical situation is generally described by the following, Lindblad-form master equation (in the interaction picture) for the pair of qubits [3]:

\begin{equation}
{d\rho_{12}\over dt} = -\imath [H_{LS}, \rho_{12}] + \sum_{\nu,i,j}\left(
\gamma_{ij}(\nu) \left[ A_j(\nu)\rho_{12} A_i^{\dag}(\nu) - {1\over 2} \{A_i^{\dag}(\nu)A_j(\nu),\rho_{12}\}\right]
\right)
\end{equation}

Our task in this section is, starting from equations (1)-(3), to derive the explicit forms for the  damping functions $\gamma_{ij}(\nu)$ and the
Lindblad operators $A_i(\nu)$; for simplicity, but without loss of generality, we ignore the Lamb shift term $H_{LS}$.

The interaction picture is defined by the self-Hamiltonian

\begin{equation}
H_{\circ} = H_{1\circ} + H_{2\circ}  +  H_{12} + H_{E_1\circ}   = {\omega \over 2} \sigma_{1z} + {\omega \over 2} \sigma_{2z} + {\beta \over 4} \sigma_{1z} \otimes \sigma_{2z} + H_{E_1\circ},
\end{equation}

\noindent where the environmental self-Hamiltonian: $H_{E_1\circ} =  \int_0^{\nu_{max}} d\nu a^{\dag}_{\nu} a_{\nu}$ with the maximal  frequency $\nu_{max}$.
The alternative choice [3,13] of the interaction picture without the $H_{12}$ term in eq.(5) reduces the considerations to the weak qubits interaction, $\beta \ll 1$, which is a special case of our considerations.

From eq.(5) it readily follow the energy eigenvalues and eigenprojectors for the qubits system:

\begin{eqnarray}
&\nonumber& E_1 = \omega + {\beta \over 4} , \quad P_1 = \vert + +\rangle\langle + +\vert,
\\&&
E_2 =  - {\beta \over 4}, \quad P_2 = \vert + -\rangle \langle + -\vert + \vert - +\rangle \langle - +\vert
 \nonumber,
\\&&
 E_3 = -\omega + {\beta \over 4}, \quad P_3=\vert - -\rangle\langle - -\vert.
\end{eqnarray}

\noindent where $\vert mn\rangle \equiv \vert m\rangle \vert n\rangle, m,n\in\{+,-\}$ and $\sigma_{z}\vert\pm\rangle=\pm\vert\pm\rangle$.
From eq.(6) follows the set of the values for the parameter $\nu$ in eq.(4): $\{0,\nu_1=E_1 - E_2 = \omega + \beta/2, \nu_2 = E_1 - E_3 = 2\omega,
\nu_3 = E_2 - E_3 = \omega - \beta/2\}$.

The general expressions for the Lindblad operators [3]:

\begin{equation}
A_i(\nu) = P_n A_i P_m, \quad \nu = E_m - E_n
\end{equation}

\noindent where $A_i =S_{1x}$, cf. eq.(3), while:

\begin{equation}
\gamma_{kl}(\nu) = 2\pi tr \left[B_k(\nu)B_l \rho_{E_1th} \right].
\end{equation}

\noindent In eq.(8) [3]: $B_k(\nu) = \int_{-a}^a dt exp(-\imath \nu t) B_k(t)$
for the $B_k(t)$ representing the interaction-picture form of $B_k$. In eq.(3) there is only one such operator, $B_{E_1}$.

The desired master equation can be shortly presented as:

\begin{equation}
{d\rho_{12} \over dt} =   - \alpha^2 \left(\sum_{i=1}^2
\left(\gamma_i
\mathcal{A}_i + \gamma^{-}_i  \mathcal{A}^{\dag}_i
\right)
\right)[\rho_{12}],
\end{equation}

\noindent where  $\alpha$  presents the weak coupling constant of the  system-environment interaction; if $\gamma\equiv\gamma(\nu)$, then $\gamma^{-}\equiv\gamma(-\nu)$, with all superoperators satisfying:

\begin{equation}
\mathcal{A}[\rho_{12}] = A\rho_{12}A^{\dag} - {1\over 2}\{A^{\dag} A, \rho_{12}\}, \quad \mathcal{A}^{\dag}[\rho] = A^{\dag}\rho_{12}A - {1\over 2}\{A A^{\dag}, \rho_{12}\}.
\end{equation}

\noindent  From (6) and (7) straightforwardly follow the non-zero Lindblad operators in eq.(10):

\begin{eqnarray}
&\nonumber&
A_1 ={1\over 8} \sigma_{1-} \otimes(I_2 - \sigma_{2z}), \nonumber
\\&&
A_2 ={1\over 8} \sigma_{1-} \otimes(I_2 + \sigma_{2z}).
\end{eqnarray}

Calculation of the damping functions $\gamma_{kl}(\nu)$
is straightforward; technical details are presented in [7].
For completeness, we provide a few main steps.

Due to the only one term in eq.(3), the general expression eq.(8) reduces to:

\begin{equation}
   \gamma_{xx}(\nu)=2\pi h(\nu) \int_0^{\nu_{max}}d\nu' h(\nu') tr (a_{\nu} (a_{\nu'}+a^{\dag}_{\nu'})\rho_\mathrm{th}).
  \end{equation}

With the use of  expressions for the thermal averages, i.e. when the environment is in thermal equilibrium [14]:

\begin{equation}
\overline{a_{\nu'}a_{\nu}}=0=\overline{a_{\nu'}^{\dag}a_{\nu}^{\dag}},
\end{equation}

\begin{equation}
\overline{a_{\nu'}a_{\nu}^{\dag}}=\delta(\nu'-\nu)(1+\bar{n}(\nu'))
\end{equation}

\noindent and

\begin{equation}
\overline{a_{\nu'}^{\dag}a_{\nu}}=-\delta(\nu'-\nu)\bar{n}(\nu'),
\end{equation}

\noindent follow the expressions  for $\gamma_{xx}(\nu)$ and $\gamma_{xx}(-\nu)$:

\begin{equation}
\gamma_{xx}(\nu)=2\pi J(\nu)(1+\bar{n}(\nu)), \quad \gamma_{xx}(-\nu)=2\pi J(\nu)\bar{n}(\nu),
\end{equation}

\noindent where $\nu$ takes the above distinguished values and the average number of bosons in thermal state $\bar{n}(\nu)=
(e^{-\nu/T}-1)^{-1}$. We choose the standard Ohmic spectral density $J(\nu)=\alpha\nu e^{-\nu/\nu_c}$ with the cutoff $\nu_c$.

In the high temperature limit, which we are concerned with, $\bar{n}(\nu)\gg 1$ and therefore
$\gamma(\nu) \approx \gamma(-\nu)=2\pi J(\nu)\bar{n}(\nu)$, which reduces the list of the damping functions to only two of them (for fixed $\nu$):
$\gamma_1\equiv \gamma(\nu_1)$ and $\gamma_2\equiv\gamma(\nu_3)$.

Substitution of eqs. (10) and (11) in eq.(9), after simple calculation, gives the following
relation of the damping functions to the Lindblad operators appearing in eq.(9):

\begin{eqnarray}
&\nonumber&
\gamma_1\equiv 4\pi J(\omega+\beta/2)\bar{n}(\omega+\beta/2) , \quad A_2,  A^{\dag}_2, \nonumber
\\&&
\gamma_2\equiv 4\pi J(\omega-\beta/2)\bar{n}(\omega-\beta/2) ,   \quad A_1, A^{\dag}_1,
\end{eqnarray}

\noindent that completes the master equation (9).

{\bf 3. Derivation of the Kraus operators}

\bigskip

In this section we derive the interaction-picture Kraus operators for the qubits dynamics described by the master equation (9).
We use a method recently developed in Ref.[8].
\bigskip

{\bf 3.1 A brief overview of the method}

\bigskip

In Ref. [8], the authors developed a general procedure
for deriving a Kraus decomposition from a known master
equation and vice versa, regarding the finite-dimensional
quantum systems. The only assumption is that the master
equation is local in time.

If the dynamical map for the process eq.(9) is formally presented as:

\begin{equation}
\rho_{12}(t) = \Phi_t[\rho_{12}(0)],
\end{equation}

\noindent and the master equation eq.(9) is shortly presented as:

\begin{equation}
{d\rho_{12}(t)\over dt} = \Lambda_t[\rho_{12}(0)],
\end{equation}

\noindent then (for the time-independent superoperator $\Lambda$),
the following matrix relation is fulfilled:

\begin{equation}
F = e^{Lt}.
\end{equation}

\noindent The matrices $F=(F_{ij})$ and $L=(L_{ij})$ are well defined for the finite-dimensional systems, and representations for the map $\Phi_t$ and for the superoperator
$\Lambda$, respectively, in a chosen orthonormalized basis $\{G_i\}$ of hermitian operators acting on the system's Hilbert state space.

Introducing the so-called Choi matrix [8]:

\begin{equation}
S_{nm} = \sum_{r,s} F_{rs} tr\left(G_rG_nG_sG_m\right)
\end{equation}

\noindent and its non-negative (real) eigenvalues $d_i$, follow the desired Kraus operators:

\begin{equation}
K_i = \sum_j \sqrt{d_j} u_{ij} G_j,
\end{equation}

\noindent where the unitary matrix $U=(u_{ij})$ diagonalizes the Choi matrix $S=(S_{nm})$.

That is, the procedure provides a Kraus form of the process:

\begin{equation}
\rho_{12}(t) = \sum_i K_i(t) \rho_{12}(0) K^{\dag}_i(t).
\end{equation}

\noindent The trace preservation implies:

\begin{equation}
\sum_i K^{\dag}_i(t) K_i(t)=I, \forall{t}.
\end{equation}

\bigskip

{\bf 3.2 The Kraus operators for eq.(9)}

\bigskip
Distilled from Section 3.1, the procedure for derivation of the Kraus operator is as follows: First, from the master equation eq.(9),
the $L$ matrix is derived. Then due to eq.(20), the $F$ matrix follows that, in accordance with eq.(21), provides the Choi matrix $S$. Finally, diagonalization of the Choi matrix gives rise to the Kraus operators, eq.(22).

We proceed by first obtaining the $\Lambda$ operator in the standard  representation of the $\sigma_{1i} \otimes \sigma_{2j}/2, i,j=0,1,2,3$, operators; $\sigma_{\circ} = I/\sqrt{2}$, while for $i>0$, the $\sigma_i/\sqrt{2}$s
represent the standard (normalized) Pauli operators. The nonzero marix elements are as follows: $-8(\gamma_1+\gamma_2) = L_{2,2}=L_{3,3}=L_{5,5}=L_{6,6}=L_{8,8}=L_{9,9}=L_{10,10}=L_{11,11}=L_{12,12}=L_{13,13}=L_{14,14}=L_{15,15}$, $-b(\gamma_1-\gamma_2)=L_{2,14}=L_{3,15}=L_{14,2}=L_{15,3}$, $-16(\gamma_1+\gamma_2)=L_{4,4}=L_{16,16}$ and $-16(\gamma_1-\gamma_2)=L_{4,16}=L_{16,4}$.

From the $L$ matrix easily follows the $F$ matrix with the following non-zero entries: $1=F_{1,1}=F_{7,7}$,
$(\exp(-16t\gamma_1)+\exp(-16t\gamma_2))/2=F_{2,2}=F_{3,3}=F_{14,14}=F_{15,15}$,
$(\exp(-16t\gamma_1)-\exp(-16t\gamma_2))/2=F_{2,14}=F_{3,15}=F_{14,2}=F_{15,3}$,
$\exp(-8t(\gamma_1+\gamma_2))=F_{5,5}=F_{6,6}=F_{8,8}=F_{9,9}=F_{10,10}=F_{11,11}=F_{12,12}=F_{13,13}$,
$(\exp(-32t\gamma_1)+\exp(-32t\gamma_2)/2 = F_{4,4}=F_{16,16}$ and
$(\exp(-32t\gamma_1)-\exp(-32t\gamma_2)/2 = F_{4,16}=F_{16,4}$.

The rest of the calculation is also straightforward but rather involved. Hence we just give the final
expressions for the non-zero Kraus operators:

\begin{equation}
K_1={\sqrt{1-e^{-32t\gamma_2}}\over 2}
\left(\begin{array}{cccc}
0&0&0 &0\\
0&0&0 &\imath\\
0&0&0 &0\\
0&-\imath &0 &0
\end{array}\right),
K_2={\sqrt{1-e^{-32t\gamma_2}}\over 2}
\left(\begin{array}{cccc}
0&0&0 &0\\
0&0&0 &-1\\
0&0&0 &0\\
0&-1 &0 &0
\end{array}\right),
\end{equation}

\begin{equation}
K_3={\sqrt{1-e^{-32t\gamma_1}}\over 2}
\left(\begin{array}{cccc}
0&0&1 &0\\
0&0&0 &0\\
1&0&0 &0\\
0&0 &0 &0
\end{array}\right),
K_4={\sqrt{1-e^{-32t\gamma_1}}\over 2}
\left(\begin{array}{cccc}
0&0&-\imath &0\\
0&0&0& 0\\
\imath&0&0 &0\\
0&0 &0 &0
\end{array}\right),
\end{equation}

\begin{equation}
K_5={1-e^{-16t\gamma_2}\over 2}
\left(\begin{array}{cccc}
0&0&0 &0\\
0&-1&0 &0\\
0&0&0 &0\\
0&0 &0 &1
\end{array}\right),
K_6={1-e^{-16t\gamma_1}\over 2}
\left(\begin{array}{cccc}
1&0&0 &0\\
0&0&0 &0\\
0&0&-1 &0\\
0&0 &0 &0
\end{array}\right).
\end{equation}

The  last two matrices are diagonal and are given by the respective entries, where we use the following notation: $\tau = (\gamma_1+\gamma_2)t$,
$W=(\gamma_1-\gamma_2)/(\gamma_1+\gamma_2)$.
For $K_7$: $K^7_{1,1}=K^7_{3,3}=A
(-8e^{32\tau}+2e^{24\tau}\sinh(16W\tau) + 4e^{32\tau}\sinh(8W\tau)+B)$;
$K^7_{2,2} = K^7_{4,4} = A(-8e^{32\tau}-2e^{24\tau}\sinh(16W\tau) - 4e^{32\tau}\sinh(8W\tau)+B)$.

\noindent For the $K_8$ matrix:
$K^8_{1,1}=K^8_{3,3}=-A'(8e^{32\tau}-2e^{24\tau}\sinh(16W\tau) - 4e^{32\tau}\sinh(8W\tau)+B)$;
$K^8_{2,2}=K^8_{4,4}=-A'(8e^{32\tau}+2e^{24\tau}\sinh(16W\tau) + 4e^{32\tau}\sinh(8W\tau)+B)$.

The notation for $K_7$ and $K_8$:

\begin{equation}
A=\sqrt{2e^{-16\tau} + 2e^{-32\tau}\cosh(16W\tau) + 4e^{-24\tau} \cosh(8W\tau) - e^{-56\tau}B
 \over
 16(2e^{16\tau}\sinh(16W\tau)+4e^{24\tau}\sinh(8W\tau))^2 + 16e^{-16\tau}(B-8e^{32\tau})^2 }
\end{equation}

\begin{equation}
A'=\sqrt{2e^{-16\tau} + 2e^{-32\tau}\cosh(16W\tau) + 4e^{-24\tau} \cosh(8W\tau) + e^{-56\tau}B
 \over
 16(2e^{16\tau}\sinh(16W\tau)+4e^{24\tau}\sinh(8W\tau))^2 + 16e^{-16\tau}(B+8e^{32\tau})^2 }
\end{equation}

\noindent and

\begin{eqnarray}
&\nonumber& B^2=
2e^{48\tau}(28e^{16\tau}-1) + 2e^{48\tau} \cosh(32W\tau) - 8e^{56\tau}\cosh(8W\tau)
\\&&
+ 8e^{64\tau} \cosh(16W\tau) + 8e^{280\tau} \cosh(120W\tau).
\end{eqnarray}

From equations (25)-(30) follows the completeness relation eq.(24) for every instant of time $t$.
Hermiticity of the Kraus operators, $K^{\dag}_i = K_i, \forall{i}$, implies the unital character
of the map $\Phi_t$ described by eq.(9), that is, $\sum_i K_i I K_i = I$; equivalently $\Phi_t[I]=I$, see eq.(18), i.e. $\Lambda_t[I]=0$, see eq. (19).

In the initial instant of time $t=0$ ($\tau=0$), $B=8$, and therefore $A=0$ while $A'=1/16$.
Therefore $K_i(0)=0, i=1,2,...,7$, while $K_8(0)=-I$, thus satisfying the initial condition,
$\rho_{12}(0)=K_8(0) \rho_{12}(0) K_8(0)$. From eqs.(25)-(27) it follows the time independence of the Kraus operators $K_i,
i=1,2,...6$ in the asymptotic limit. That is

\begin{equation}
\lim_{t\to\infty}\sum_{i=1}^6 K_i(t) K_i(t)= I/4.
\end{equation}

\noindent Then, due to the completeness relation eq.(24), it follows the dominant contribution from the $K_{7}$ and $K_8$  operators:

\begin{equation}
\lim_{t\to\infty}\sum_{i=7}^8 K_i(t) K_i(t)= 3I/4.
\end{equation}

Of course, this does not imply existence of a stationary state, which is defined in the Schr\" odinger picture, in which the Kraus operators are defined as:

\begin{equation}
U^{(12)}_{\circ}(t) K_i(t), i=1,2,...,8,
\end{equation}

\noindent where $U^{(12)}_{\circ}(t)=\exp(-\imath t (H_{\circ}-H_{E_1\circ}))$ and $H_{\circ}$ is defined by eq.(5).

\bigskip

{\bf 3.3 Entanglement sudden death for the pair of qubits}

\bigskip

As an application of the results of Section 3.2, we investigate entanglement dynamics for the pair of qubits described by the master equation (9).
We assume the initial maximally entangled state for the pair of qubits, $(\vert +  -\rangle + \vert - +\rangle)/\sqrt{2}$, and choose the following set of the values
for the parameters appearing in eq.(17):
$\omega =0.1, \alpha = 0.02, T =100,  \nu_c =100 $. As a measure of quantum entanglement we use the standard and well-studied measure of {\it concurrence} [15]

\begin{equation}
\mathcal{C}(\rho(t)) = \max\{0, \Lambda(t)\},
\end{equation}

\noindent where $\Lambda(t) = \sqrt{\lambda_1(t)} - \sqrt{\lambda_2(t)}- \sqrt{\lambda_3(t)}-\sqrt{\lambda_4(t)}$ with the eigenvalues
$\lambda_1 > \lambda_2>\lambda_3>\lambda_4$ of

\begin{equation}
\rho(t) (\sigma_{1y}\otimes\sigma_{2y})\rho^{\ast}(t)(\sigma_{1y}\otimes\sigma_{2y})
\end{equation}

\noindent and "$\ast$" denoting the complex-numbers conjugate. The density matrix $\rho(t)$ in eq.(35) is the $1+2$-system's state eq.(23) for the Kraus operators given by eq.(33).

In Fig.1 we can see dependence of $\mathcal{C}$ on both, time $t$ and the strength of the inter-qubits interaction $\beta$.
Expectably, we observe dynamical decrease of $\mathcal{C}$ and hence of the entanglement in the qubits $1+2$ system for every chosen value of $\beta$.
Fig.2 depicts slower decrease of $\mathcal{C}$  for larger $\beta$, for every instant of time $t$. That is, the largest (negative) value of $\Lambda(t)$
increases with the increase of $\beta$. Physically, Figures 1 and 2 reveal
detrimental influence of the environment on the initial entanglement for the pair of qubits; the stronger inter-qubits interaction (the larger $\beta$)
the qubits more efficiently "hold together". Our results reveal the phenomenon of the so-called entanglement sudden death [16]:
instead of the expected smooth, {\it asymptotic} approach to $\mathcal{C}=0$, we obtain dynamical change of the concurrence from the initial $\mathcal{C}(0)=1$ to the final $\mathcal{C}(t)=0$
value for the {\it finite} time interval $t$.

\begin{figure}[!ht]
\centering
 \includegraphics[width=0.35\textwidth]{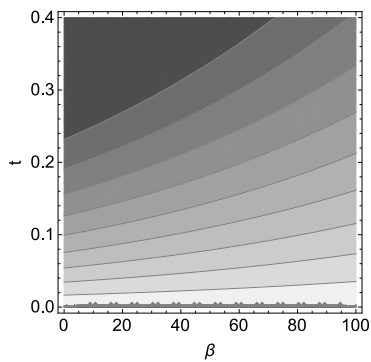}

\caption{Concurrence presented as a function of both time $t$ and the inter-qubit interaction $\beta$.
The   larger values are shown lighter, while  the parameters  are given in the body  text.}
\end{figure}

\begin{figure}[!ht]
\centering
 \includegraphics[width=0.35\textwidth]{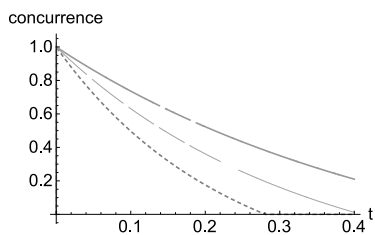}

\caption{Concurrence dynamics  with $\beta =0, 50, 100$ presented respectively by the dashed, thin and thick lines.}
\end{figure}

\bigskip

{\bf 4. Discussion}

\bigskip

Procedure presented in this paper is universal, in that it formally equally applies to arbitrary physical situation for a pair of qubits that is subjected to a
completely positive dynamical process. For both qubits monitored by their respective environments (or by a common environment),
the technicalities are much more involved but without any conceptual or methodological obstacles or open issues. The use of the Kraus operators is a straightforward
application of the matrix (linear algebra) calculus.

Tracing out a qubit from eq.(9) universally leads to a master equation for the other qubit. However, this procedure is trivial only
for the initial tensor-product state $\rho_1\otimes\rho_2$ for the pair of qubits. A qubit can be regarded as a part of the (extended) environment
of the other qubit. This gives rise to alternative bipartitions of the total system.

Tracing out that includes the qubit $2$ (the qubit $1$) regards
the $1+E'_1\equiv 1+(E_1+2)$ (the $2+E'_2\equiv 2+(1+E_1)$) bipartition. Only if there is not correlations in the initial state of the qubits,
the bipartition of the total system, $(1+2)+E_1$, which is assumed in Section 2 and leads to eq.(4), is interchangeable with the alternative
bipartitions of the total system: $(1+E_1)+2$ and $1+(E_1+2)$.  The presence of the initial correlations in the qubits $1+2$ system implies
initial correlations [17] in both $1+E'_1$ and $2+E'_2$ bipartitions, and hence [3]
non-complete-positivity of the processes for the individual qubits; for this reason, in general, independent derivations of the master equations for the individual qubits are required [18,19].
That is, the initial correlations in the qubits system breaks the symmetry between the different bipartitions of the total system.
For this reason we only regarded the bipartition $(1+2)+E_1$ and eq.(4) as the  reliable basis for the analysis of the qubits dynamics--the absence of the initial
correlations in this bipartition is supposed from the very start, $\rho_{12}\otimes\rho_{E_1}$, Section 2.

For a pair of qubits, every Kraus operator can be written $K_k = \sum_{i,j} c^k_{ij} A_{1i}\otimes B_{2j}$; an example is given by eq.(11). Then the state for the pair of qubits reads:

\begin{equation}
\rho_{12}(t) = \sum_{k,i,i',j,j'} c^k_{ij} c^{k\ast}_{i'j'} A_{1i}\otimes B_{2j} \rho_{12}(0) A^{\dag}_{1i'}\otimes B^{\dag}_{2j'},
\end{equation}

\noindent which after the tracing out the qubit $2$ (the qubit $1$) gives a master equation for the qubit $1$ (for the qubit $2$). As emphasized above, only for the initial
tensor-product state $\rho_{12} = \rho_1\otimes\rho_2$, one can obtain a Kraus form for a single-qubit dynamics--that guarantees [1] complete positivity of the
process the qubit is subjected to.  Then eq.(36) gives, e.g.:

\begin{equation}
\rho_1 = tr_2 \rho_{12} = \sum_{i,i'} A_{1i} \rho_1 A^{\dag}_{1i'} \left(tr_2 \sum_{k} (\sum_j c^k_{ij} B_2j) \rho_2 (\sum_j c^k_{ij} B_2j)^\dag\right).
\end{equation}

\noindent Since the terms

\begin{equation}
b_{ii'} = tr_2 \sum_{k} (\sum_j c^k_{ij} B_{2j}) \rho_2 (\sum_j c^k_{i'j} B_{2j})^\dag \equiv \sum_k tr_2 B_{2ik} \rho_2 B^{\dag}_{2i'k}
\end{equation}

\noindent  constitute a positive semi-definite matrix, diagonalization of the $(b_{ii'})$ matrix,
$b_{ii'} = b_k w_{ik} w^{\ast}_{i'k}$, $b_k\ge 0, \forall{k}$, gives rise to the Kraus operators for the qubit $1$:

\begin{equation}
K^1_k=\sum_i \sqrt{b_k} w_{ik} A_{1i},
\end{equation}

\noindent where the unitary matrix $(w_{ik})$ diagonalizes the $(b_{ii'})$ matrix.
As emphasized above, this procedure, and complete positivity of the qubit's dynamics, breaks for any kind of the initial correlations in the $1+2$ system [3, 13].
To this end, a more detailed  analysis with an emphasis on the subtleties regarding the very concept
of complete positivity [3, 13, 20] will be presented elsewhere.

An important extension of the standard procedure for Markovian dynamics is considered for a pair of weekly interacting damped harmonic oscillators [3, 13]; in our considerations,
this is the $\beta \ll 1$ case. Then the interaction term $H_{12}$,
eq.(2), may be regarded as a perturbation and an alternative interaction picture can be used by omitting $H_{12}$  in eq.(5). Then $H_{12}$ appears  in the commutator term of the master equation with the Lindblad operators, which are obtained for the case $\beta=0$, cf. eq.(B.1) in Ref. [13].
Nevertheless. this is just a special case of our considerations. Due to the commutation $[H_{1\circ}+H_{2\circ},H_{12}]=0$,
the Lindblad operators eq.(11) are the same for both $\beta \neq 0$ and $\beta =0$. Of interest is the interaction picture state defined as
$\tilde \rho_{\circ}(t)=U^{\dag}_{\circ}\rho(t)U_{\circ}$, where $U_{\circ}=\exp(-\imath t(H_{1\circ}+H_{2\circ}+H_{E_1\circ}))$; $\rho(t)$ is in the Schr\" odinger-picture. The exact interaction picture state
$\tilde \rho(t) = U^{\dag}\rho(t)U$, where $U(t)=\exp(-\imath t(H_{1\circ}+H_{2\circ}+H_{12}+H_{E_1\circ}))$. Keeping the terms of the first order in (small) $\beta$, the approximation $U
\approx U_{\circ}(I-\imath t H_{12})$ easily gives:

\begin{equation}
\tilde\rho_{\circ}(t) \approx \tilde\rho(t) - \imath t [H_{12},\tilde\rho_{\circ}(t)].
\end{equation}

Taking the time derivative of eq.(40)  easily follows, in the new interaction picture:

\begin{equation}
{d\tilde\rho_{\circ} \over dt} \approx -\imath [{\tilde H_{1E_1},\tilde\rho(t)}] - \imath[H_{12}, \tilde\rho_{\circ}(t)]\approx -\imath [\tilde H_{1E_1} + H_{12}, \tilde\rho_{\circ}(t)],
\end{equation}

\noindent which is eq.(B.1) in Ref. [13]; $H_{12}$ is of the same form for both pictures. We also note that, while eq.(40) is universal (for sufficiently small $\beta$),
in general, the Lindblad operators for the two interaction pictures are not identical; nevertheless, even in the more general cases, our approach
(approximation of the above unitary operator $U(t)$),  in the zeroth order of the time-independent perturbation, leads to eq.(B.1) of Ref. [13].

\bigskip

{\bf 5. Conclusion}

\bigskip

Derivation of the Kraus operator-sum from the microscopic Hamiltonian model may be technically involved.
This may be the reason behind the lack of the explicit forms of the Kraus operators for most of the non-single-qubit processes.
Nevertheless, the use of the Kraus form of the dynamical map is often technically simple. Our results aim at reducing this gap and exhibiting the
technical advantage of the use of a Kraus form for certain basic tasks in open systems and quantum information theory contexts.

\bigskip

{\bf References}

\bigskip

[1] K. Kraus, States, effects and operations, fundamental notions of
quantum theory, Springer (1983)

[2] H.P. Breuer, F. Petruccione, The theory of open quantum systems
(Clarendon, Oxford, 2002)

[3] \' A. Rivas, S. F. Huelga, Open quantum systems--an introduction
(Springer Briefs in Physics) (2012)

[4] J. Jekni\' c-Dugi\' c, M. Arsenijevi\' c, M. Dugi\' c,  Proc. R. Soc. A {\bf 470} (2014) 20140283.

[5] J. Jekni\' c-Dugi\' c, M. Arsenijevi\' c, M. Dugi\' c, Proc. R. Soc. A {\bf 472} (2016) 20160041.

[6] M. A. Nielsen, I. L. Chuang, Quantum Computation and Quantum
Information (Cambridge Univ. Press, Cambridge, 2000)

[7] M. Arsenijevi\' c, J. Jekni\' c-Dugi\' c, M. Dugi\' c, Braz. J. Phys. (2017) {\bf 47}, 339–349

[8] E. Andersson, J.D. Cresser, M.J.W. Hall, J. Mod. Opt. 54, 1695
(2007)

[9] L. E. Ballentine, Phys. Rev. A {\bf 43}, 9 (1991)

[10] B. Vacchini,  Int. J. Theor. Phys. {\bf 44}  (2005) 1011-1021

[11] S. J. Yun et al, J. Phys. B, At. Mol. Opt. Phys. \textbf{48}
(2015) 075501

[12] K. M. Fonseca Romero, R. Lo Franco, Phys. Scr. 86, 065004
(2012)

[13]  \' A. Rivas, A. Douglas, K. Plato, S. F Huelga, M. B Plenio,
New Journal of Physics {\bf 12} (2010) 113032

[14] W. H. Louisell, Quantum Statistical Properties of Radiation (New York, Wiley, 1973) p. 182

[15] W.K. Wootters, Phys. Rev. Lett. {\bf 80}, 2245 (1998)

[16] T. Yu, J.H. Eberly, Phys. Rev. Lett. {\bf 93}, 140404 (2004)

[17] M. Arsenijevi\' c, J. Jekni\' c-Dugi\' c, M. Dugi\' c, Chin. Phys. B {\bf 22},  020302 (2013)

[18] M. Arsenijevi\' c, J. Jekni\' c-Dugi\' c, D. Todorovi\' c, M. Dugi\' c, 2015, Entanglement Relativity in the Foundations of The Open Quantum Systems Theory, in: New Research on Quantum Entanglement, Ed. Lori Watson, Nova Science Publishers, 2015, pp. 99-116

[19] R. E. Kastner, J. Jekni\' c-Dugi\' c, G. Jaroszkiewicz, eds. Quantum Structures. Classical Emergence
from the Quantum Level, World Scientific, Singapore, 2017.

[20] A. Brodutch, A. Datta, K. Modi, \' A. Rivas, C. A. Rodríguez-Rosario, Phys. Rev. A {\bf 87}, 042301 (2013)

\end{document}